\documentclass[twocolumn,showpacs]{revtex4}
\usepackage{epsfig,color}

\def\$K^-${$K^-$}
\def\K+{$K^+$}
\def\AG{$A\cdot$GeV}
\begin{document}

\title{Observation of different azimuthal emission patterns for
$K^+$ and of $K^-$ mesons in Heavy Ion Collisions at 1-2~\AG}

\author{F.~Uhlig$^b$, A.~F\"orster$^{b,+}$, I.~B\"ottcher$^d$,
M.~D\c{e}bowski$^{e,f}$, F.~Dohrmann$^f$, E.~Grosse$^{f,g}$,
P.~Koczo\'n$^a$, B.~Kohlmeyer$^d$, F.~Laue$^{a,*}$, M.~Menzel$^d$,
L.~Naumann$^f$, H.~Oeschler$^b$, W.~Scheinast$^f$, E.~Schwab$^a$,
P.~Senger$^a$, Y.~Shin$^c$, H.~Str\"obele$^c$, C.~Sturm$^{b,a}$,
G.~Sur\'owka$^{a,e}$,
A.~Wagner$^f$, W.~Walu\'s$^e$\\
(KaoS Collaboration)\\
$^a$ Gesellschaft f\"ur Schwerionenforschung, D-64220 Darmstadt, Germany\\
$^b$ Technische Universit\"at Darmstadt, D-64289 Darmstadt, Germany\\
$^c$ Johann Wolfgang Goethe-Universit\"at, D-60325 Frankfurt am Main, Germany\\
$^d$ Phillips Universit\"at, D-35037  Marburg, Germany\\
$^e$ Jagiellonian University, PL-30059 Krak\'ow, Poland\\
$^f$ Forschungszentrum Rossendorf, D-01314 Dresden, Germany \\
$^g$ Technische Universit\"at Dresden, D-01062 Dresden, Germany\\
$^*$ Present address: Brookhaven National Laboratory,  USA \\
$^+$ Present address: CERN, CH-1211 Geneva 23, Switzerland }

\begin{abstract}
Azimuthal distributions of  $\pi^+$, $K^+$
and $K^-$ mesons have been measured in Au+Au reactions at 1.5~\AG
~and Ni+Ni reactions at 1.93 \AG. In semi-central collisions  at
midrapidity, $\pi^+$ and $K^+$ mesons are emitted preferentially
perpendicular to the reaction plane in both collision systems.
In contrast for $K^-$ mesons in Ni+Ni reactions an in-plane elliptic
flow was observed for the first time at these incident energies.

\end{abstract}


version of \today

\maketitle

Relativistic heavy ion collisions provide a unique opportunity to
study both the behavior of nuclear matter at high densities as
well as the properties of hadrons in dense nuclear matter. In
particular, strange mesons are considered to be sensitive to
in-medium modifications. Theory predicts a repulsive
$K^+N$ potential and an attractive $K^-N$ potential in dense
matter \cite{schaf}. It is suggested that the latter effect leads
to a condensate of $K^-$ mesons in the interior of the neutron
stars, causing dramatic consequences for the neutron star
stability \cite{brobet}.

Microscopic transport calculations simulating heavy-ion collisions
predict  measurable consequences of the in-medium modifications of
strange mesons. The  $KN$ potentials reduce the $K^+$ yield and
enhance the $K^-$ yield, resulting in an increase of the
$K^-/K^+$ ratio.

First experimental evidence for in-medium
modifications of $K^-$ mesons in dense nuclear matter was the
observation that the $K^-/K^+$ ratio was enhanced in Ni+Ni
collisions as compared to nucleon-nucleon collisions \cite{Barth,
Menzel}.
A large $K^{-}/K^{+}$ ratio was also found in C+C \cite{laue} and
in Au+Au collisions \cite{AF}.  In heavy-ion collisions,  however,
strangeness-exchange reactions like $\pi\Lambda\rightarrow K^- N$
contribute significantly to the production of $K^-$ mesons. This
process, although taken into account by transport calculations,
reduces the sensitivity of the $K^-$ meson yield to the in-medium
$KN$ potential~\cite{ko84,Hart02,HO,AF}.

Another observable affected by in-medium effects is the azimuthal
emission pattern of $K^+$ and $K^-$ mesons in heavy-ion collisions.
$K^+$ and $K^-$ mesons experience different
potentials in nuclear matter: While the scalar potential acts
attractively on both kaon species, the vector potential repels
$K^+$ mesons and attracts $K^-$ mesons.
For the $K^+$ mesons these two contributions almost
cancel each other leading to a small repulsive $K^+N$ interaction.
The superposition of the two attractive interactions results in a
strongly attractive potential for $K^-$ mesons~\cite{schaf}.

A repulsive $K^+N$ potential would repel the $K^+$ mesons from the
bulk of the nucleons and therefore cause an preferred out-of-plane
emission of the $K^+$ mesons at midrapidity and a directed flow
opposite to the nucleons at target and projectile rapidity. These
effects were found in experiments \cite{shin, crochet} and
interpreted as evidence for a repulsive $K^+N$ potential \cite{Li,
Fuchs2}.

The propagation of $K^-$ mesons in nuclear matter is governed by a
large $K^-p$ cross section of up to 100 mb which is dominated by
inelastic scattering via the strangeness-exchange reaction $\pi
Y\leftrightarrow K^- N$ with $Y=\Lambda, \Sigma$. Therefore, one
would expect a pronounced azimuthal anisotropy of the $K^-$ meson
emission in heavy-ion collisions due to the interaction of $K^-$
mesons with spectator matter. However, when taking into account
the strongly attractive in-medium $K^-N$ potential, transport
calculations predict an almost isotropic  azimuthal emission
pattern at midrapidity \cite{Fuchs2}.

In this Letter we present experimental data on the azimuthal
distributions of both $K^+$ and $K^-$ mesons in nucleus-nucleus
collisions. We have measured two systems: Ni+Ni at 1.93 \AG ~(both
for $K^+$ and $K^-$ mesons) and Au+Au at 1.5 \AG ~(only $K^+$
mesons). For comparison also results from an analysis of $\pi^+$
mesons are given. An azimuthal distribution of $K^-$ mesons
emitted in heavy-ion collisions at subthreshold beam energies is
shown for the first time.

The experiments were performed with the Kaon Spectrometer (KaoS)
at the heavy-ion synchrotron (SIS) at GSI in Darmstadt
\cite{senger}, using an Au beam of 1.5 $A\cdot$GeV impinging on an
Au target (0.96 g/cm$^2$) and a Ni beam of 1.93 $A\cdot$GeV  on a
Ni target (0.68 g/cm$^2$). The particles were identified using the
momentum and time-of-flight information of the magnetic
spectrometer, and two hodoscopes were used for event
characterization~\cite{Brill}. The Large Angle Hodoscope is used
to derive the centrality of the collision from the multiplicity of
charged particles measured in the polar angle range 12$^{\circ} <
\theta_{lab} < 48^{\circ}$. The orientation of the event plane was
reconstructed from the azimuthal emission angle of the charged
projectile spectators using the transverse momentum method
\cite{daniel}. These particles were identified (up to $Z$ = 8) by
their energy loss and their time of flight as measured with the
Small Angle Hodoscope located about 7~m downstream from the target
covering polar angles between 0.5$^\circ$ and 11$^\circ$. The
resolution in the determination of reaction plane~\cite{Brill} is
$\langle\Delta\Phi^2\rangle^{1/2}=37^\circ$ for the Au-system and
$\langle\Delta\Phi^2\rangle^{1/2}=61^\circ$ for the Ni-system.

\begin{figure}
\epsfig{file=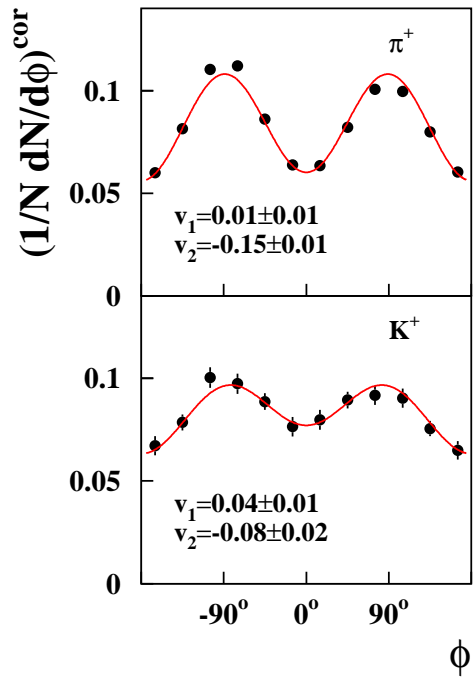,width=7.5cm} \caption{Azimuthal
distribution of $\pi^+$ and $K^+$ mesons for semi-central Au+Au
collisions at 1.5 $A\cdot$GeV. The data are corrected for the
resolution of the reaction plane and refer to impact parameters of
5.9 fm $< b <$ 10.2 fm, rapidities of 0.3 $< y/y_{beam} <$ 0.7 and
momenta of 0.2 GeV/c $< p_t <$ 0.8 GeV/c. The lines are fits with
function~(\ref{v1_v2}) resulting in the values for $v_1$ and $v_2$
as given in the figure.} \label{Au_midrap}
\end{figure}

Figure~\ref{Au_midrap} shows the azimuthal distributions of $K^+$
and $\pi^+$ mesons for semi-central Au+Au collisions at 1.5 \AG.
The distributions are corrected for the angular resolution of the
reaction plane determination~\cite{Brill}. The data are fitted
using the first two components of a Fourier series
\begin{equation}
\frac{dN}{d\Phi}\sim 2\, v_1 \cos(\phi) \, + \, 2\, v_2 \cos (2\phi)
\label{v1_v2}
\end{equation}
resulting in values for $v_1$ and $v_2$, as given in the figures
together with the statistical errors. The determination of the
coefficient $v_1$ is subject to an additional systematic error of
0.04.

Both $\pi^+$ and $K^+$ mesons exhibit a pronounced enhancement  at
$\phi = \pm 90^o$, i.e.~perpendicular to the reaction plane. For
$\pi^+$ mesons this effect can be interpreted as rescattering and
absorption at the spectator fragments. The data are in agreement
with previous observations~\cite{shin,Brill}.

\begin{figure}
\epsfig{file=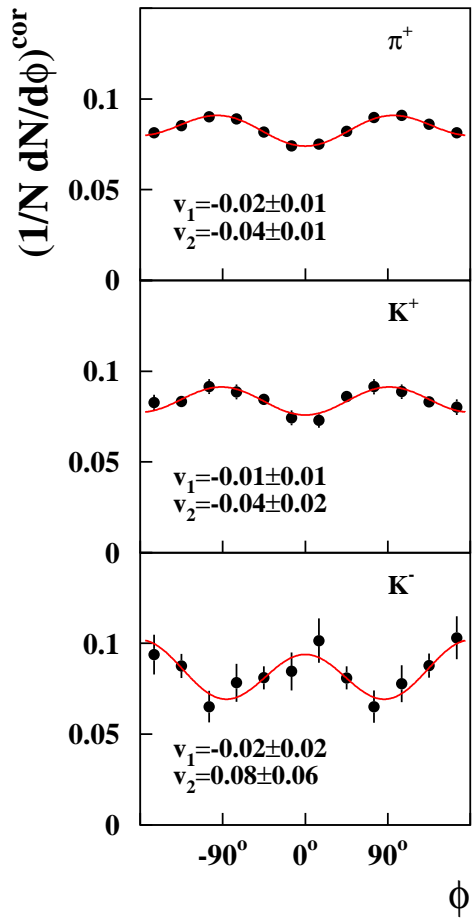,width=7.5cm} \caption{Azimuthal
distribution of $\pi^+$, $K^+$ and $K^-$ mesons for semi-central
Ni+Ni collisions at 1.93 $A\cdot$GeV. The data are corrected for
the resolution of the reaction plane and
 refer to impact parameters of 3.8 fm$ < b <$ 6.5 fm, rapidities of
0.3 $< y/y_{beam} <$ 0.7 and momenta of  0.2 GeV/c $< p_t <$ 0.8 GeV/c.
The lines are fits with function~(\ref{v1_v2}) resulting in the values
for $v_1$ and $v_2$ as given in the figure.}
\label{Ni_midrap}
\end{figure}

The study of Ni+Ni collisions was performed at a higher
incident energy of 1.93 \AG. The resulting higher production cross
section for $K^-$ mesons provides an opportunity to study both charged kaon
species. The data are shown in Fig.~\ref{Ni_midrap} along with $\pi^+$
mesons for semi-central  Ni+Ni collisions. Both $\pi^+$ and $K^+$
mesons follow the same trend already observed in Au+Au collisions.
The values for $v_2$ are
smaller than in Au+Au as one might expect for the smaller system.
In contrast to the $\pi^+$ and $K^+$ mesons, the $K^-$ mesons
show an in-plane enhancement.

This ``positive'' (in-plane) elliptic flow of particles is
observed for the first time in heavy-ion collisions at SIS
energies. In contrast to this observation one would expect a
preferential out-of-plane emission (negative elliptic flow) of
$K^-$ mesons due to their large absorption cross section in spectator
matter.

A depletion of the expected out-of-plane emission pattern of $K^-$ mesons
might be due to the fact that they are produced via
strangeness-exchange reactions. This causes a delay in the
freeze-out of the $K^-$ mesons~\cite{AF,Hart02,HO}, and, hence, a
reduced shadowing effect by the spectator fragments which have
moved further away. The observed in-plane emission of $K^-$ mesons,
however, cannot be easily explained with this scenario.

\begin{figure}
\epsfig{file=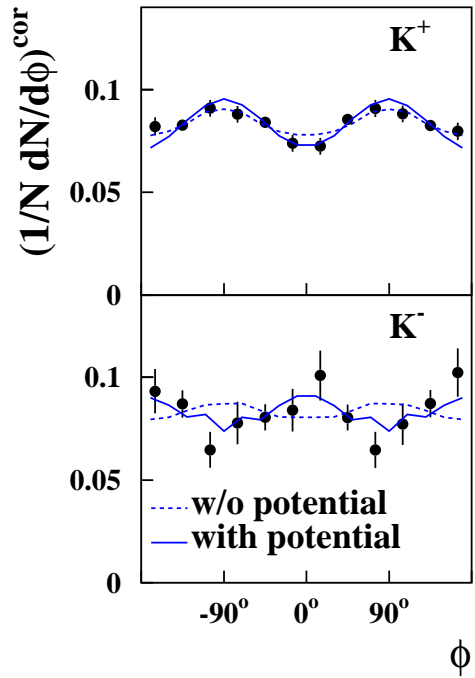,width=7.5cm}
 \caption{Comparison of the data from Ni+Ni reactions at 1.93 \AG ~with IQMD
  model calculations \cite{hart01}. The data are
corrected for the resolution of the reaction plane and
 refer to impact parameters of 3.8 fm$ < b <$ 6.5 fm, rapidities of
0.3 $< y/y_{beam} <$ 0.7 and momenta of  0.2 GeV/c $< p_t <$ 0.8 GeV/c.}
\label{iqmd_calculation}
\end{figure}

In order to quantitatively explain the $K^+$ and $K^-$ meson
azimuthal distributions we compare in Fig.~\ref{iqmd_calculation}
our data to recent results of the Isospin Quantum Molecular
Dynamics (IQMD) model \cite{hart01}. This transport calculation
takes into account both the space-time evolution of the reaction
system and the in-medium properties of the strange mesons.  The
dashed and solid lines represent results of calculations without
and with in-medium potentials, respectively.  In the case of the
$K^+$ mesons (upper panel of Fig.~\ref{iqmd_calculation}) the
effect of the repulsive $K^+N$ potential is small in this model. A
large fraction of the observed out-of-plane enhancement, in
contrast to other models \cite{Fuchs2,Li,Mishra}, is caused by the
scattering of $K^+$ mesons. Another transport code (HSD)
\cite{Mishra} predicts a dominant influence of the potential on
the emission pattern of the $K^+$ mesons. In the system Au+Au at 1
\AG ~where both size and life time of the fireball are larger than
in the Ni+Ni case, the effect of the repulsive $K^+N$ potential
was studied using the RBUU code and was found to be very important
\cite{shin,Li,Fuchs2}.

In the lower part of Fig.~\ref{iqmd_calculation} we compare a
calculation without (dashed) and with (solid) $K^-N$ potential.
When neglecting the $K^-N$ potential, the calculation predicts a
weak in-plane elliptic flow caused by shadowing. This effect is
rather small because of the late emission of $K^-$ mesons. When
taking into account the attractive in-medium $K^-N$ potential the
model is able to describe the experimental in-plane elliptic flow
pattern much better. Model calculations with the HSD
code~\cite{Mishra} predict a flat azimuthal distribution both with
and without a $K^-N$ potential and, hence cannot  explain the
observed in-plane flow of $K^-$ mesons.

In summary, we have measured the azimuthal emission patterns of
$\pi$ and K mesons in heavy-ion collisions at threshold beam
energies. We found a pronounced out-of-plane emission (negative
elliptic flow) for the $K^+$ mesons confirming previous results.
We presented new data on the azimuthal angle distribution of $K^-$
mesons which exhibit a positive (in-plane) elliptic flow pattern,
which is in contrast to all other measured particles exhibiting a
preferred out-of-plane emission.

This observation can be explained by a transport model (IQMD) by
a late emission of $K^-$ mesons and assuming an attractive in-medium
$K^-N$ potential. Hence, the distribution of strange mesons in
space and their multiplicity which has been used so far, are
independent probes to extract information on in-medium properties
at high densities.

\end{document}